\documentclass[11pt]{article}
\usepackage{moriond,epsfig}
\usepackage{amsmath}
\usepackage{float}
\usepackage{floatflt}

\usepackage{subfig}

\newcommand{\IP}{I\!P}

\def\be{\begin{equation}}
\def\ee{\end{equation}}
\def\bea{\begin{eqnarray}}
\def\eea{\end{eqnarray}}

\begin{document}
\vspace*{4cm}
\title{HADRONIC FINAL STATES AND DIFFRACTION AT HERA }

\author{ I.A. RUBINSKIY (ON BEHALF OF THE H1 AND ZEUS COLLABORATIONS) }

\address{DESY, Notkestrasse 85,\\
Hamburg, Germany}

\maketitle\abstracts{
Recent results from the $ep$ (electron-proton) collider HERA are presented.
Jet production and diffraction type reactions allow tests of predictions of Quantum Chromodynamics (QCD).
Jet measurements are used to improve the combined H1-ZEUS parton density fits
 by adding an additional constraint on the gluon distribution in the proton and, 
at the same time, providing information on $\alpha_s$. 
Measurements of inclusive diffraction and diffractive dijet data are discussed in the context of factorisation in diffraction.
Diffractive $\Upsilon(1S)$ production was studied to extract the $t$-slope of the process, 
which was found to be consistent with the other exclusive vector meson data and can be interpreted in terms of the gluonic radius of the proton.
}

\section{Introduction}
At the $ep$ collider HERA, electrons of 27.6 GeV were collided with 920 (820) GeV protons. 
HERA effectively provided photon-proton collisions at a centre-of-mass energy $W$, up to its maximum value, set by the centre-of-mass energy of the electron-proton system $\sqrt{s}$ = 310 GeV. 
The photon virtuality $Q^2$ ranged up to several thousand GeV$^2$.
The reaction phase space can be divided into photoproduction ($\gamma p$), $Q^2 \approx$ 0 GeV$^2$, and deep-inelastic scattering (DIS), where $Q^2>$ 1 GeV$^2$. 
From the first DIS experiments on the proton it is known that the proton has a structure, which can not be calculated from the first principles of QCD. 
The inclusive double differential DIS cross section can be expressed in terms of the proton structure functions $F_2(x,Q^2)$ and $F_L(x,Q^2)$,
where $x$ in lowest order corresponds to the fraction of proton momentum carried by the struck parton. 
The structure functions are strongly related to the parton density functions (PDF).
The Q$^2$ dependence of the PDFs is described by  evolution equations~\cite{DGLAP}.
At low values of $x$ ($x < 10^{-2}$) the gluon PDFs significantly exceed the quark contribution. 



At certain conditions the struck parton in its hadronization process can form a jet, a highly collimated group of particles.
Both the DIS and $\gamma p$ jet production can be used to constrain the proton and photon PDFs and measure the strong coupling constant, $\alpha_s$.




Diffractive reactions, when the scattered proton stays intact, at HERA constitute about 10$\%$ of the total cross section.
The corresponding diffractive PDFs can be defined and, due to their unversality, applied to predict diffractive processes cross section at the Tevatron or LHC experiments. 
Exclusive diffractive vector meson production provides a tool for estimating the transverse distribution of gluons in the proton.

In this contribution, recent results from the H1 and ZEUS collaborations on jet production and diffractive reactions are presented and their impact on proton PDFs and $\alpha_s$ measurements are discussed.

\section{Hadronic Final States and Diffraction at HERA}\label{sec:hera}

\subsection{Jets, $\alpha_s$ and PDFs}\label{subsec:jets}

Cross section predictions can be factorised into a perturbatively calculable QCD part (pQCD), 
where a "hard" scale for a series expansion in $\alpha_s$ can be defined, and a "soft" part, which consists of the proton PDFs and hadronization scheme of the partons produced in the reaction.
Perturbative calculations lead to partonic final states which are not directly accessible to the measurement. 
The observed hadrons or dense groups of hadrons, called jets, are the results of the fragmentation of coloured partons. 
Typical "hard" scales in jet production are the $Q^2$ or a the jet transverse energy, $E_T$.

\begin{figure}[h]
\centering
\subfloat[]{\includegraphics[width=2.0in]{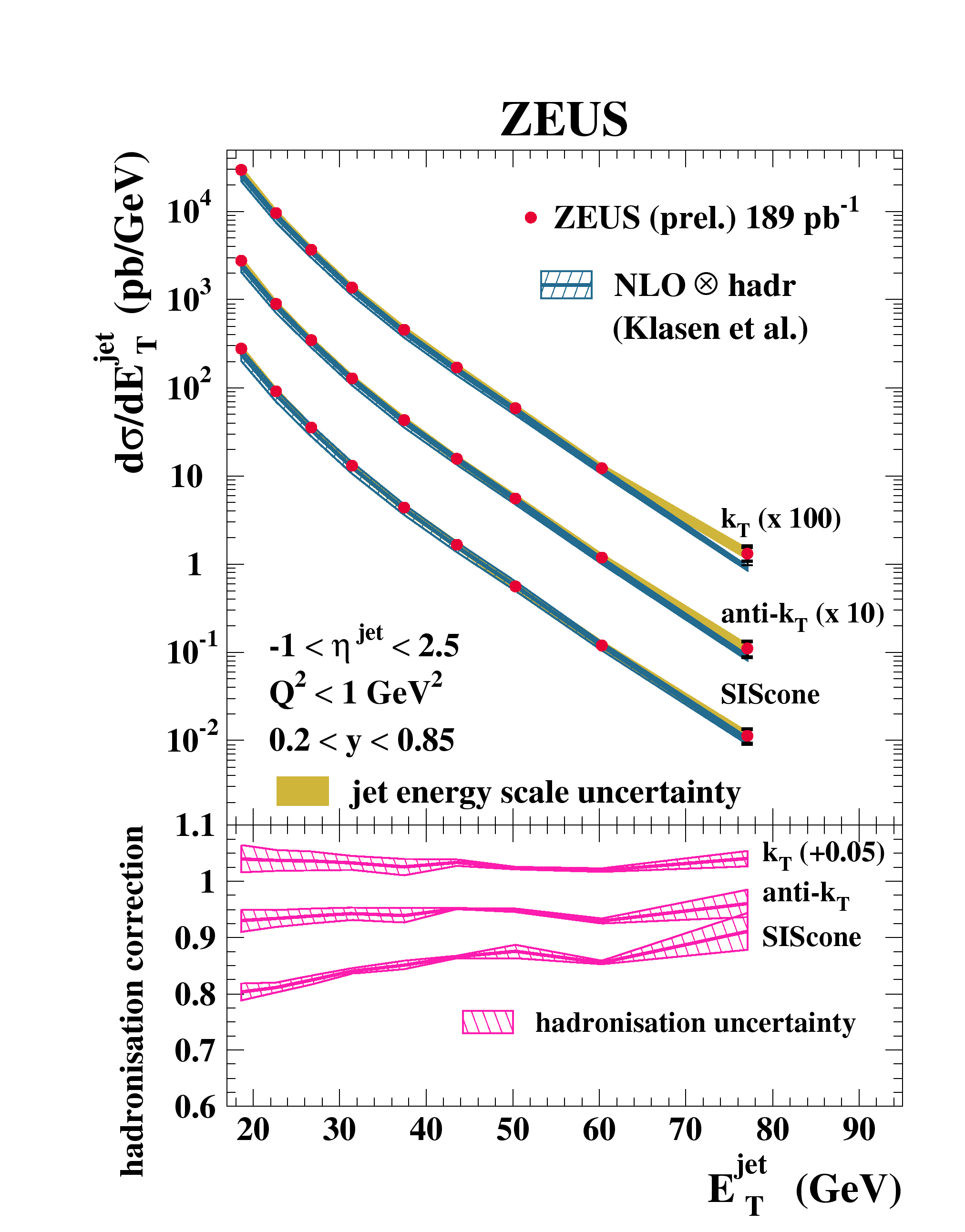}}
\subfloat[]{\includegraphics[width=2.1in]{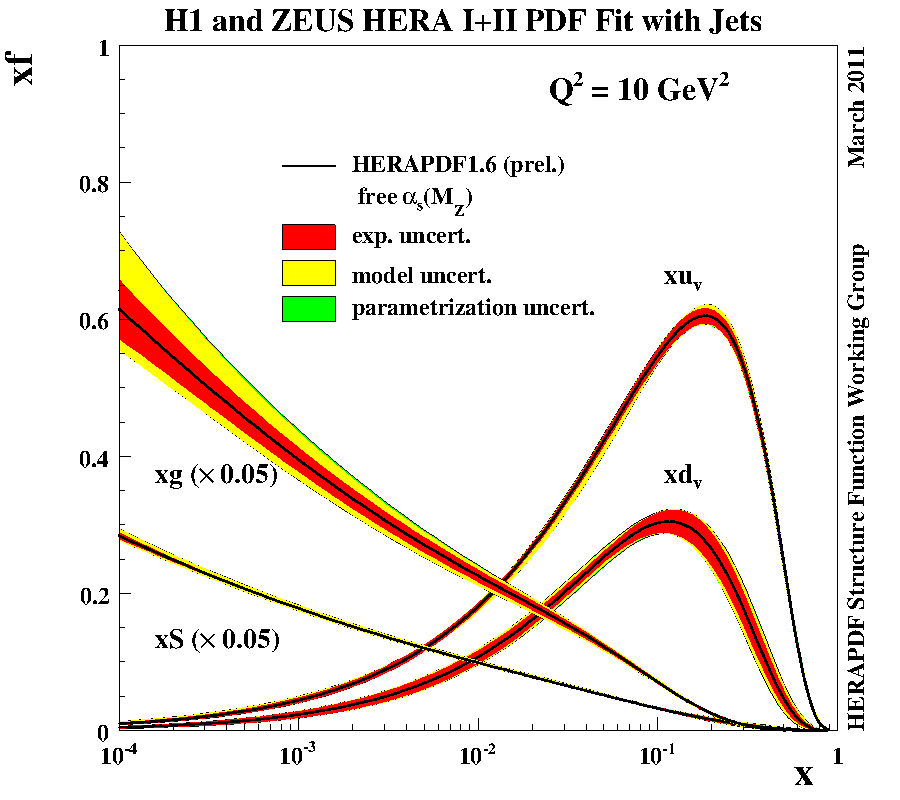}}
\subfloat[]{\includegraphics[width=2.1in]{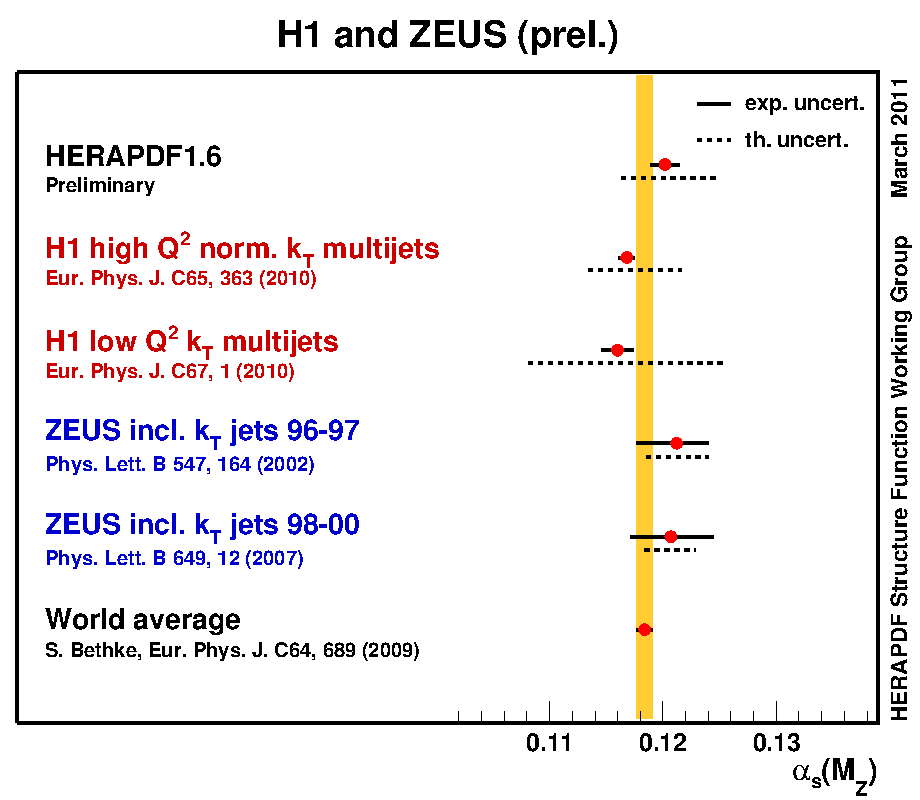}}
\caption{
a) comparison of three jet clustering algorithms,
b) HERAPDF 1.6 proton PDFs (with the jet data) and 
the c) HERAPDF 1.6 fit $\alpha_s$  measurement (the data subsets used in the fit and their $\alpha_s$ measurements are shown).
}
\label{fig:jets}
\end{figure}

Several algorithms are used to identify jets at hadron level, like $k_T$, or newly developed, like anti-$k_T$ and SIScone.
Since the details of each jet algorithm are different (e.d. radius, combination scheme), the hadronization corrections are different for different algorithms.
Benchmarking of the jet algorithms has been performed by ZEUS both for $\gamma p$~\cite{PHP-kt-antikt-ZEUS}, and DIS~\cite{DIS-kt-antikt-ZEUS} jet production. 
The recently developed infrared- and collinear-safe jet algorithms, anti-kt~\cite{CacciariSalamSoyez} and SIScone~\cite{SalamSoyez}, have been shown to be compatible with the previously widely used $k_T$ algorithm. Theoretical uncertainties on PDFs and $\alpha_s$ are very similar for all three algorithms, corrections for the terms beyond the next-to-leading order (NLO) and hadronization are very similar for $k_T$ and anti-$k_T$, but larger for SIScone (Fig.~\ref{fig:jets}a).



The jet production cross sections have the potential to constrain the proton and photon PDFs~\cite{proton-gamma-PDFs-ZEUS}.
It has been demonstrated in the combined H1-ZEUS HERAPDF 1.6 fit~\cite{combined-PDF-alfas} that by adding jet data 
to the inclusive DIS cross sections the fit procedure does not depend on external information on the strong coupling constant $\alpha_s$ and can provide an independent $\alpha_s$ measurement (Fig.~\ref{fig:jets}b,c). 


\subsection{Diffraction, $F_2^{D}$, $F_L^{D}$, dijets, $\Upsilon(1S)$ }\label{subsec:diff}
About 10$\%$ of all electron-proton collisions at HERA are diffractive. 
Diffractive reactions are characterized  by the exchange of a colorless object with the quantum  numbers of the vacuum (often called Pomeron). 
Experimentally diffractive reactions can be selected by a large rapidity gap (LRG) between the reaction products or by explicit tagging of the proton with a proton spectrometer system.
Due to the factorisation theorem for diffractive DIS~\cite{Collins98} it is possible to define diffractive PDFs of the proton.
The diffractive inclusive double differential DIS cross section can be written as: 

\begin{equation}
\label{ln:eq:diffxsec}
\frac{d^4\sigma^{ep}_D(x_{\IP},\beta,t,Q^2,)}{dx_{\IP}d\beta dtdQ^2}=\frac{4\pi\alpha^2_{em}}{xQ^4}Y_{+}[F_2^D(x_{\IP},\beta,t,Q^2)-\frac{y^2}{Y_{+}}F_L^D(x_{\IP},\beta,t,Q^2)],
\end{equation}

with a so called reduced cross section: 

\begin{equation}
\label{ln:eq:reddiffxsec}
\sigma^{D(4)}(x_{\IP},\beta,t,Q^2)=F_2^D(x_{\IP},\beta,t,Q^2)-\frac{y^2}{Y_{+}}F_L^D(x_{\IP},\beta,t,Q^2),
\end{equation}

where $x_{\IP}$ is the \IP$ $omeron momentum fraction, 
$\beta$ - the momentum fraction of the hard scattered parton,  and
$t$ - the squared 4-momentum transfer at the proton interaction vertex.

\begin{figure}[h]
\centering
\subfloat[]{\includegraphics[width=6.2in]{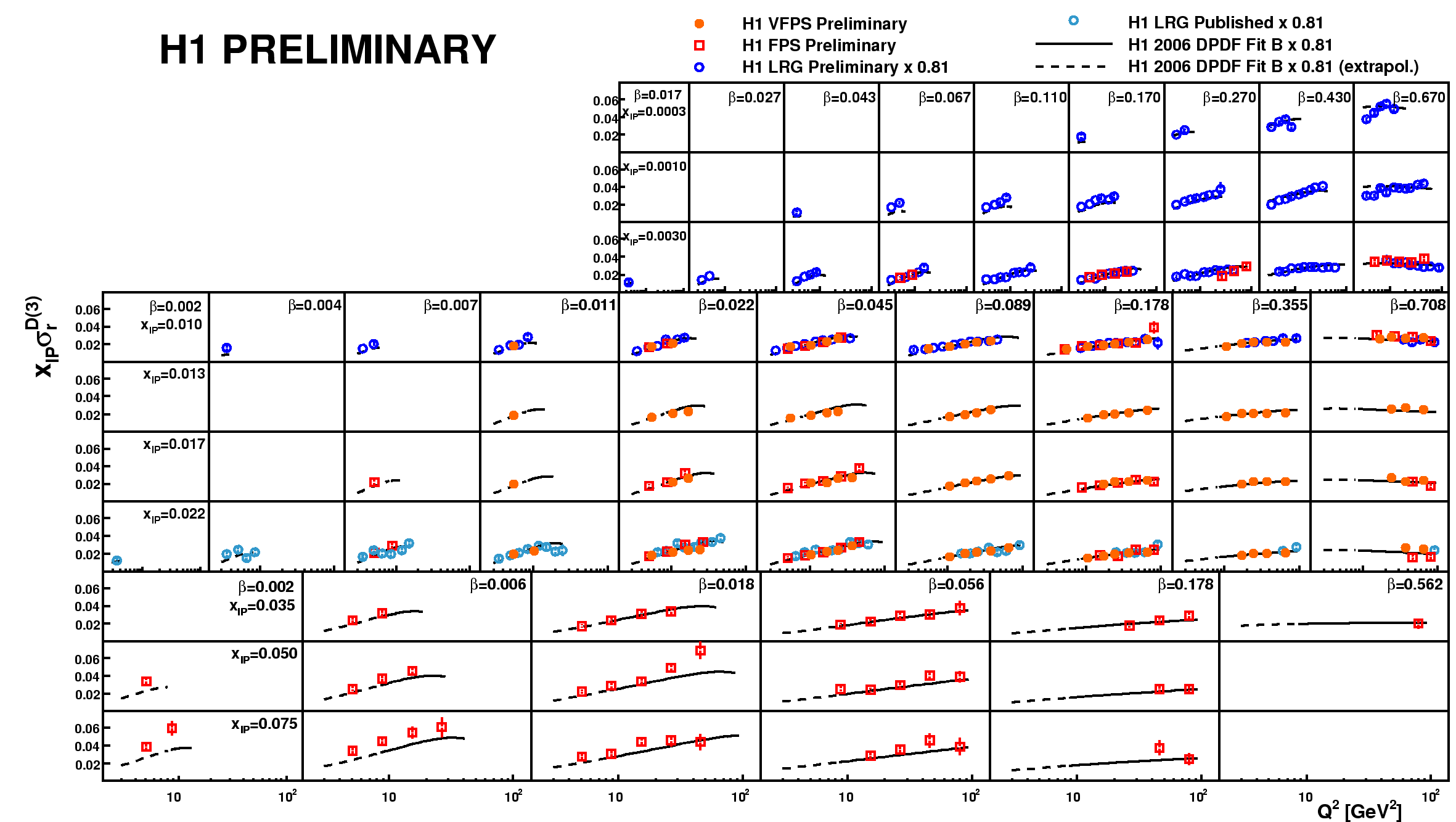}} \linebreak
\subfloat[]{\includegraphics[width=2.2in]{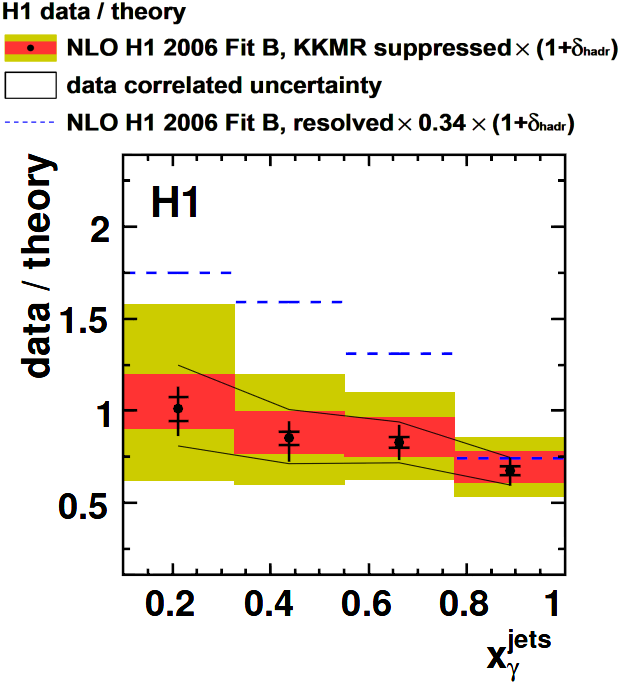}}
\subfloat[]{\includegraphics[width=4.0in]{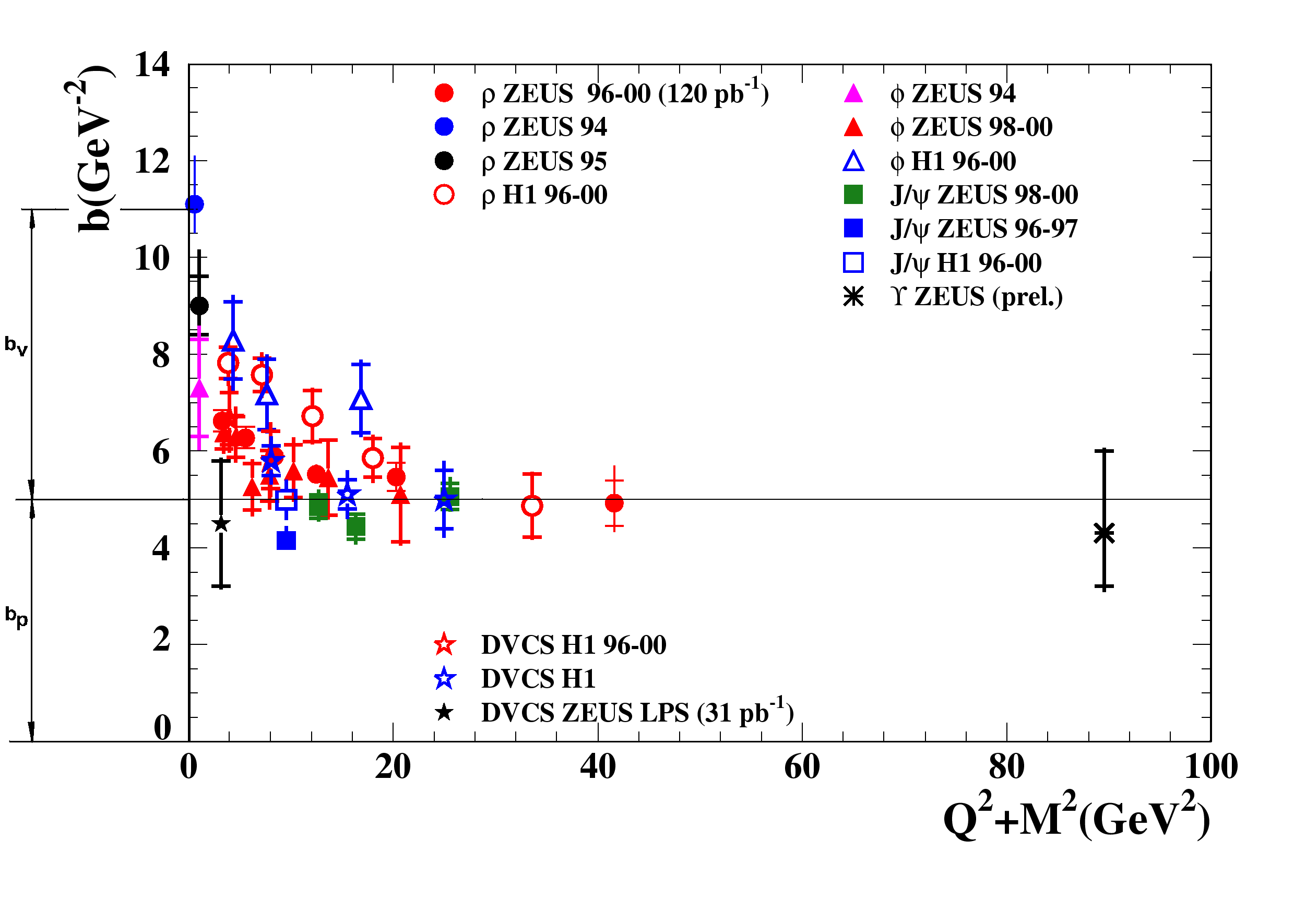}}
\caption{
a) Reduced cross section measured by proton spectrometers, FPS and VFPS, and LRG method.
b) The ratio of the measured to the predicted diffractive dijet cross sections  as function of the photon momentum fraction $x_{\gamma}$ after the "gap suppression" correction implementation.
The NLO QCD predicted cross section was done based on the FR framework and the H1 2006 Fit B DPDF set, corrected for hadronisation effects. 
The FR theoretical prediction for resolved photons is modified by applying the scale factors from the KKMR model for point-like interactions (‘KKMR suppressed’)$^{18}$
or for hadron-like interactions (‘resolved ×0.34’)$^{19}$.
c) Diffractive process $t$-slope, $b$, for light and heavy vector mesons measured at different $Q^2$ values. 
The value $b$ is expected to be a sum of two components coming from the vector meson, $b_V$, and the proton, $b_p$. 
}
\label{fig:f2d3}
\end{figure}

The H1 proton spectrometers, the FPS (Forward Proton Spectrometer) and the VFPS (Very Forward Proton Spectrometer), were used for proton tagging and measurement of the  $F_2^D(4)$~\cite{H1FPSF2D4}. 
Since the LRG method does not provide the $t$ measurement one can use only the $\sigma^{D(3)}$ for comparison with proton spectrometer method measurements.
The two different methods cover a vast region in the phase space and agree well in the acceptance overlap~\cite{H1prelim-10-011,H1prelim-10-014,DESY-10-095} (Fig.~\ref{fig:f2d3}a).

The longitudinal structure function $F_L^D$ contribution becomes non-negligible only at high $y$. 
 It has direct sensitivity to the  gluon density function. 
The measurement~\cite{H1prelim-10-017} is consistent with the  gluon density determined from the scaling violations of the F$_2^D$ structure function and from diffractive jet production.

The diffractive PDFs (DPDF), if known, allow cross section calculations for other DDIS observables. 
The calculations for DDIS at HERA have been performed with very good agreement for the jet~\cite{H10012051}
 and heavy quark production measurements~\cite{H10610076}. 
However, DPDF-based predictions for hard diffractive processes such as dijet production 
in $p\bar{p}$ scattering fail by around an order of magnitude to describe the data~\cite{CDF5043,KK0908.2531}.
The issues of DPDFs applicability and possible "rapidity gap survival probability" were studied at HERA in diffractive dijet photoproduction~\cite{DESY-10-043,H1prelim-10-013}.
In leading order the virtual photon with zero virtuality, $Q^2\approx 0$, can be viewed as either "direct" or "resolved".
In direct photon processes the photon  enters the hard interaction directly, 
while in resolved processes it enters the hard interaction via its partonic structure and  has a point-like and a hadron-like components.
There are both theoretical and experimental arguments that the "resolved"-photon contribution can be suppressed by a significant gap suppression factor. 
For the hadron-like component the "gap suppression" factor has been estimated to be $\sim$ 0.34~\cite{GAPSUP23,GAPSUP24}, 
and the point-like component suppressed by a factor 0.7-0.8 depending on the jet $E^{jet}_T$~\cite{GAPSUP24,GAPSUP25}.
It has been shown that the introduction of the "resolved" component suppression factors improves the data to theory agreement~\cite{DESY-10-043}(Fig.~\ref{fig:f2d3}b)
at low x$_\gamma$ where  resolved processes are expected to dominate.



Exclusive vector meson (VM) production at HERA can be studied in terms of the color dipole model. 
The photon-proton collision can be treated as a $q\bar{q}$ - proton interaction.
The $t$-dependence of the exclusive vector meson production can be related to the transverse size of the interaction region, which is set by the gluon distribution.
At higher values of the VM mass, $M_V$, or of $Q^2$, the $q\bar{q}$ pair transverse size vanishes, so the proton gluonic transverse size can be determined.  
As it is seen in Fig.~\ref{fig:f2d3}c the $t$-dependence slope flattens already after $M_V^2+Q^2>10$ GeV$^2$ at $b\approx$ 5 GeV$^{-2}$. 
This value can be roughly translated into $r_{gluons}\approx$ 0.5 $fm$ which is smaller then the electromagnetic radius of the proton, $r_{em}\approx 0.8$ $fm$.
A new measurement of the $\Upsilon(1S)$ diffractive production~\cite{ZEUS-prel-10-020} extends the $M_V^2+Q^2$ tested range by a factor 2 and agrees well with the other VM measurements.

\section{Summary}

A brief overview of recent results from the HERA collider published by the H1 and ZEUS collaborations is given.
The measurements provide information on the parton structure of the proton and precise measurement of the strong coupling constant $\alpha_s$. 
High energy diffraction keeps being a very interesting field for QCD studies.
Further understanding of rapidity gap suppression mechanism is required for the ongoing and future collider experiment programs.






\end{document}